\documentclass{article}
\usepackage{amsmath,amssymb,amsfonts}
\begin{document}

\par 
\LARGE 
\noindent 
{\bf Spectroscopy of an AdS Reissner-Nordstr\o m black hole} \\
\par 
\normalsize 

\noindent {\bf Claudio Dappiaggi$^{1,2,a}$}, {\bf Simona Raschi$^{1}$},\\
\par
\small
\noindent $^1$ Dipartimento di Fisica Nucleare e Teorica, Universit\`a di Pavia
via A.Bassi 6 I-27100 Pavia, Italy.

\noindent $^2$ Istituto Nazionale di Fisica Nucleare  
Sezione di Pavia, via A.Bassi 6 I-27100 Pavia, Italy.\smallskip

\noindent $^a$  claudio.dappiaggi@pv.infn.it\\
 \normalsize

\small 
\noindent {\bf Abstract}. 
{In the framework of black hole spectroscopy, we extend the results obtained for a charged black hole 
in an asymptotically flat spacetime to the scenario with non vanishing negative
cosmological constant. In particular, exploiting Hamiltonian techniques, we 
construct the area spectrum for an AdS Reissner-Nordstr\o m black hole.}

\normalsize
\newpage 

\section{Introduction}
The road leading to an explicit formulation of a full quantum theory of gravity
necessarily requires a complete comprehension of the role of extreme gravitating
objects such as black holes. In particular a key question concerns the 
characterization and the construction of their Fock space of states in a quantum 
version of general relativity. 
An answer would grant us the chance to study the spectrum of operators acting on
the Fock space and, consequently, the quantum numbers characterizing the 
expectation values. Furthermore, this project would
ultimately put to test the long standing conjecture according to which 
a discrete spectrum is associated to the area operator of the event horizon.

In this connection, a landmark has been set by Bekenstein (see \cite{Bekenstein2}
and \cite{Bekenstein3}) who suggested that the area of the event horizon
of a black hole behaves as an adiabatic invariant of the associated dynamical
system. With this hypothesis and assuming that the standard rules of quantum mechanics
could also be applied in a strong gravitational field, it is possible to appeal 
to the so-called \emph{Ehrenfest principle}. It grants that a quantum operator with a discrete spectrum
corresponds to any classical ``adiabatic'' observable on a phase space. The reasonableness of the
above assumption has been addressed by several authors and, in particular, it holds for a neutral non
rotating black holes \cite{Bekenstein7, Mayo}. The natural subsequent 
step consists of conjecturing that the most general form for the area spectrum is:
\begin{equation}\label{Bek1}
A= n a_0,
\end{equation}
where $n$ is an integer number and $a_0$ is a constant proportional to $l^2_p$,
the square of the Planck length. Under the same hypothesis, this result
have been generalized to charged black holes, such as 
Reissner-Nordstr\o m, with one notable exception: the role of adiabatic invariant is
not played by the area $A(M,q)$ depending on the mass $M$ and the charge $q$ but by
the difference between $A(M,q)$ and the area $A_{ext}(q)$ of the black hole in the
extremal classical configuration \cite{Barvinsky}.

The only setback in the above construction lies in the heuristic nature of
Bekenstein results which, thus, cannot provide any precise mean to rigorously justify the 
shape of the area spectrum and, in particular, to find an exact value for the
constant $a_0$ in (\ref{Bek1}).
 The are two notables approaches for making more precise and predictable the above
 arguments; the first has been proposed by Bekenstein himself \cite{Bekenstein2,
 Bekenstein3} and it is based on an
axiomatic point of view according to which one constructs a closed and linear algebra
of observables for the black hole Fock space such that the spectrum can be directly
inferred from the properties of the algebra itself. Within this framework, (\ref{Bek1})
is confirmed on a firmer ground though the specific value for the constant $a_0$
cannot be derived.\\
The second approach instead concentrates on a Hamiltonian formulation
\cite{Barvinsky} i.e., by exploiting a generalized Birkhoff theorem, the dynamic of a
spherically symmetric background, can be described on a ``reduced phase space'' where
the main variables are diffeomorphisms invariant physical observables quantities, 
such as the black hole mass and charge, and their associated conjugate momenta. Within
this framework the rationale is to select a suitable canonical transformation of
coordinates for the phase space in such a way that the Hamiltonian for the dynamical
system is similar to that of a harmonic oscillator. A rather direct quantization
procedure for the Hamiltonian itself and, when necessary, for the charge sector, allows
to directly retrieve the area spectrum. The advantage of this scheme over the axiomatic
one lies in the absence of any a priori assumption either on the shape of the spectrum
or on the existence of a particular algebra of operators. Furthermore the results
confirm Bekenstein suggestions and a specific value for $a_0$ can be directly 
inferred; it is also possible to show that ultimately both methods, the axiomatic and
the reduced phase space, are completely equivalent, either for charged either for
neutral, non rotating black holes \cite{Das2}.

The phase space approach has been applied to a wide class of specific black 
holes, the most notables being the Schwarzschild (see also \cite{Vaz}), the 
Reissner-Nordstr\o m \cite{Barvinsky} and the
Kerr black holes \cite{Gour, Gour2} though, in all these scenario, a condition of
asymptotic flatness, has been imposed. It is thus natural to ask ourselves whether it is
possible to generalize the above results when a different class of spacetimes is
chosen. Here we consider asymptotically AdS black holes, though we will not refer to
the neutral non rotating case which can be seen as a particular subcase of 
\cite{Barvinsky4} and of \cite{Barvinsky5}. On the opposite we concentrate on 
the AdS Reissner-Nordstr\o m case, we construct the area 
spectrum and we test if it is still equispaced even when the charged sector is taken
into account. In other words, we consider what is the
effect of the cosmological constant on the eigenvalues of the area operator.\\
Beside testing Bekenstein ideas in a broader setting, there are wider motivations 
to address this question since asymptotically AdS backgrounds play a central role in 
current research. In particular an answer to the above
query should be seen as a starting point to test the results of black hole
spectroscopy  from the point of view of the AdS/CFT 
correspondence \cite{ADS/CFT} and to compare them to other
approaches to area quantization \cite{Rovelli}, in particular that of quasi
normal modes \cite{Hod, Dreyer} where a comparison with Bekenstein
approach is still rather elusive at least in an asymptotically flat black hole
background (see \cite{Setare3} for a recent exhaustive review).\\ 

The outline of the paper is the following: in section 2 we review the reduced 
phase space approach and the scheme of quantization. We
will also discuss this method in the context of an asymptotically
flat Reissner-Nordstr\o m black hole. In section 3 we deal instead with the AdS
counterpart explicitly constructing the spectrum and commenting on the result.
In section 4 we draw some conclusion and we propose some possible future
lines of research. Eventually, in the appendix, we review the axiomatic approach
in connection with the method discussed in sections 2 and 3.
 
\section{The reduced phase space approach}
As we have outlined in the introduction, the remarkable arguments put forward by 
Bekenstein, in order to calculate the spectrum of the
event horizon area for a black hole, are heuristics and they are considered as
the starting point to set the \emph{black hole spectroscopy} in a more rigorous
framework which is here introduced and reviewed. 

\subsection{Construction of the area spectrum}\label{II}
The rigorous approach to black hole spectroscopy, we now discuss, is often known as 
{\em reduced phase space formulation} \cite{Barvinsky, Barvinsky4, Barvinsky5,
Barvinsky2, Barvinsky3}. Since our aim
is to ultimately apply this specific method to a charged spherically symmetric
non rotating AdS black hole, we start from a rather general framework, specializing 
the discussion in the next section to the asymptotically flat Reissner-Nordstr\o m black hole.\\
The starting point is a four dimensional Lorentzian smooth
manifold $\mathcal{M}^4$ and the Einstein-Maxwell action \cite{Wald}
\begin{equation}\label{act}
S=\int\limits_{\mathcal{M}^4}d^4x\sqrt{\left| g\right|}\left(\frac{R}{16\pi}-
\frac{1}{4\pi}F^{\mu\nu}F_{\mu\nu}\right),
\end{equation}
where $F_{\mu\nu}=A_{[\mu;\nu]}$, being $A_\mu$ the $U(1)$ electromagnetic potential.
If we introduce the usual coordinate system $(x,t,\theta,\varphi)$, a generic 
spherically symmetric metric can be written as
\begin{equation}\label{sym}
ds^2=g_{\alpha\beta}dx^\alpha dx^\beta+r^2(x_\alpha)d^2\Omega(\theta,\varphi),
\end{equation}
where $d^2\Omega(\theta,\varphi)$ is the solid angle element and where the strictly positive
valued function $r(x_\alpha)$, the radius 
of the sphere and the metric $g_{\alpha\beta}$ depend only upon the coordinates $x_\alpha\doteq\left
\{t,x\right\}$ spanning a two dimensional spacetime $\mathcal{M}^2$. Assuming that also $A_\mu$ is spherically symmetric, a direct 
substitution of (\ref{sym}) in (\ref{act}) provides, upon integration of the 
angular coordinates and under the rescaling $r\to\sqrt{2} r$, the dimensionally 
reduced action ({\em c.f.} section 2 in \cite{Grumiller} - see also \cite{Nojiri} and 
\cite{Nojiri2})
\begin{eqnarray}
\qquad S=\int\limits_{\mathcal{M}^2} d^2x\sqrt{\left|g\right|}\left[\frac{1}{2}
\left(\frac{g_{\alpha
\beta}}{2}\partial^\alpha r\partial^\beta
r+1+\frac{r^2}{2}R(g_{\alpha\beta})\right)-\frac{r^2}{4}F^{(2)\alpha\beta}
F^{(2)}_{\alpha\beta}\right],\label{act1}
\end{eqnarray}
where $R(g_{\alpha\beta})$ is the scalar curvature associated with the two dimensional 
metric $g_{\alpha\beta}$ and $F^{(2)}_{\alpha\beta}$ is the two dimensional 
electromagnetic field strength. 
For our purposes, dimensional reduction plays a key role since it allows us to 
perform a Hamiltonian analysis of the spherically symmetric Einstein-Maxwell
action along the lines of \cite{Medved}. In this last cited paper a more general class of 
dilatonic actions have been taken into account namely:
\begin{eqnarray}
\qquad S^\prime=\int\limits_{\mathcal{M}^2} d^2x\sqrt{\left| g\right|}\left[
\frac{1}{2}\left(\frac{g_{\alpha\beta}}{2}\partial^\alpha\psi\partial^\beta
\psi+V(\psi)+D(\psi)R\right)-\frac{W(\psi)}{4}F^{\alpha\beta}F_{\alpha\beta}\right]\label{act3}
\end{eqnarray}
where $R$ is the scalar curvature, $\psi$ plays the role of the dilaton field, $F_{\alpha\beta}$ is the electromagnetic field strength 
and where $V(\psi)$, $D(\psi)$ and $W(\psi)$ are generic functions.
Whenever $D(\psi)$ is at least a differentiable function of $\psi$ such
that $D(\psi)\neq 0$ and $\frac{dD(\psi)}{d\psi}\neq 0$, it is possible to cancel 
the kinetic term by means of a suitable field redefinition \cite{Grumiller}:
\begin{equation}\label{coordch}
\bar{g}_{\alpha\beta}=\Omega^2(\psi)g_{\alpha\beta},\quad\phi=D(\psi).
\end{equation}
Here $\Omega^2(\psi)$ is a conformal rescaling which, up to an irrelevant
constant, is set by
\begin{equation}\label{conformalfactor}
\Omega^2(\psi)=\exp\left(\frac{1}{2}\int\left(\frac{dD(\psi)}{d\psi}\right)^{-1}
d\psi\right). 
\end{equation}
According to (\ref{coordch}), (\ref{act3}) becomes
\begin{equation}\label{act4}
S^\prime=\int\limits_{\mathcal{M}^2} d^2x\sqrt{\left|\bar{g}\right|}
\left[\frac{1}{2}\left(\phi R+V(\phi)
\right)-\frac{W(\phi)}{4}F^{\alpha\beta}F_{\alpha\beta}\right].
\end{equation}
An important property of this action has been discussed in \cite{Louis-Martinez} and in 
\cite{Louis-Martinez2} where a Birkhoff-like theorem has been
proved. In detail, it is always possible to find a suitable local coordinate 
frame such that the metric arising from a solution of the Euler-Lagrange 
equations of (\ref{act4}) is static and it depends only upon diffeomorphism
invariant parameters. If we introduce a suitable time coordinate
$\tau$
and the spatial coordinate $\sigma\equiv \phi$, the most general solution can be written as
\begin{gather}\label{metric}
ds^2=-f(\sigma; C, q)d\tau^2+f^{-1}(\sigma; C,q)d\sigma^2,\\
f(\sigma; C,q)=-C+j(\sigma)-q^2k(\sigma),\label{metric2}\\
F_{\alpha\beta}=\frac{q}{W(\sigma)}\epsilon_{\alpha\beta},\label{metric3}
\end{gather}
where $C,q$ are two constants of integration which are respectively related to 
the $ADM$ mass and to the $U(1)$ electric charge, as we will clarify when we will deal
with the Hamiltonian formulation. The functions $j(\sigma),k(\sigma)$ instead are implicitly
defined from the equations
\begin{equation*}
V(\sigma)=\frac{dj(\sigma)}{d\sigma},\quad W^{-1}(\sigma)=\frac{dk(\sigma)}{d\sigma}. 
\end{equation*}
 
We can now discuss the above mentioned Hamiltonian analysis; as a starting point
we assume that $\mathcal{M}^2$ is locally $\mathbb{R}\times \mathcal{C}$ where 
$\mathcal{C}$ is a one dimensional spatial manifold, not necessarily closed.
A generic metric can be written as
\begin{equation*}
ds^2=e^{2\rho}\left[-N^2_1dt^2+\left(dx+N^2_2dt\right)^2\right],
\end{equation*}
where $x$ now is a local coordinate for $\mathcal{C}$ and $\rho,N_1,N_2$ are generic
functions of $(t,x)$. Plugging this metric in (\ref{act4}) and slavishly
repeating the analysis of \cite{Louis-Martinez}, we find that the momenta
associated to $N_1$, $N_2$ and to $A_0$, vanish; consequently these fields
play the role of Lagrange multipliers and, besides the dilaton $\phi$, only
$\rho$ and $A_1$ are dynamical degrees of freedom. If we introduce the conjugate
momenta $\Pi_\rho,\Pi_\phi,\Pi_{A_1}$, the full Hamiltonian is the sum of two components, the first is the 
canonical Hamiltonian $H_{can.}$ whereas the second is a surface term needed whenever $\mathcal{C}$ is non 
compact in order for the Hamilton equations to be fully consistent with the Euler-Lagrange equations i.e.:
\begin{gather}
\mathcal{F}=\rho^\prime\Pi_\rho+\phi^\prime\Pi_\phi-\Pi^\prime_\rho\sim 0,
\quad \mathcal{I}=-\Pi^\prime_{A_1}\sim 0,\label{const1}\\
\mathcal{G}=\phi^{\prime\prime}-2\phi^\prime\rho^\prime -2\Pi_\phi\Pi_\rho 
-e^{2\rho}V(\phi)+\frac{e^{2\rho}}{W(\phi)}\Pi^2_{A_1}\sim 0,\label{const2}\\
H=H_{can.}+H_{surf.}=\int\limits
dx\left[N_1\frac{\mathcal{G}}{2}+N_2\mathcal{F}+A_0\mathcal{I}\right]+H_{sur.},\label{Hamil}
\end{gather}
where $\mathcal{F},\mathcal{G}$ generate 
spacetime diffeomorphisms and $\mathcal{I}$ enforces the $U(1)$ gauge 
transformation. These are secondary constraints and the symbol $\sim 0$ implies that they are  weakly 
vanishing in the Dirac sense ({\em c.f.} chapter 12 of \cite{Dirac}). Consequently we are dealing 
with a dynamical system with six first class constraints, three primary $N_1, N_2, A_0$ and three secondary 
$\mathcal{F}, \mathcal{G}, \mathcal{I}$; according to \cite{Louis-Martinez} and to 
\cite{Louis-Martinez3}, the latter can be solved as:
\begin{equation}\label{momenta}
\Pi_\rho=Q[C,q,\rho,\phi],\quad\Pi_\phi=\frac{g[q,\rho,\phi]}{4Q[C,q,\rho,\phi]},
\quad\Pi_{A_1}=q,
\end{equation}
where $C,q$ are two constants of integration which turn out to be the same as those introduced in 
(\ref{metric2}) i.e. they are respectively (proportional to) the ADM mass and the electric charge. Moreover 
a straightforward counting shows that, to each point in $\mathcal{M}^2$, it is associated a six dimensional 
phase space and the six above mentioned first class constraints, thus there are no
propagating modes in the dynamical system described by
(\ref{act4}).

The next step lies in the key observation that $C,q$ have vanishing Poisson brackets 
with all the constraints and between them; consequently they can be written in terms of the original 
phase space coordinates as
\begin{equation}\label{laC}
C=e^{-2\rho}\left[\Pi_\rho^2-\left(\phi^\prime\right)^2\right]+j(\phi)-k(\phi)\Pi^2_{A_1},\quad q=\Pi_{A_1},
\end{equation}
where, as before, $\frac{dj(\phi)}{d\phi}=V(\phi)$ and $\frac{dk(\phi)}{d\phi}=W^{-1}(\phi)$. \\
Thus, from a direct inspection of (\ref{const1}) and of (\ref{const2}), it appears that both $C$ and $q$ are
independent from the spatial coordinate on $\mathcal{C}$, but, more importantly,
they are also the the unique independent physical observables in the Dirac 
sense \cite{Louis-Martinez}.

Let us now introduce the canonical momenta
associated to $C,q$ namely $\Pi_C,\Pi_q$ referring for an explicit expression and a 
discussion about their specific properties to section IV and VI in 
\cite{Louis-Martinez}. We simply remark that,
according to the analysis in \cite{Gegenberg},  $\Pi_q$ is related to the asymptotic choice for the 
$U(1)$ gauge whereas, if we consider an evolution between two different instants of
time, $\Pi_C$, at infinity, can be physically interpreted as the time separation. 
Furthermore the following key relationship holds \cite{Barvinsky, Louis-Martinez}:
\begin{equation}\label{variazione}
\delta\Pi_q=\frac{\Phi}{2}\delta\Pi_C+\delta\lambda,
\end{equation}
where $\delta$ refers to a variation under a change on the boundary conditions
and where $\Phi$ is the electrostatic potential calculated at the boundary under
consideration. If we take into account the above remarks and the absence of 
propagating modes, we can switch from the six dimensional phase space generated 
by $\left(\rho,\phi,A_1,\Pi_\rho,\Pi_\phi,\Pi_{A_1}\right)$ to a {\em reduced 
phase space} $\Gamma\sim\mathbb{R}^4$ generated by $\left(M,q,\Pi_M,\Pi_q\right)$.

In order to complete the analysis of the system ruled by  
(\ref{act4}), the last step consists of discussing $H_{surf.}$ in (\ref{Hamil}) i.e.
we need to impose suitable boundary conditions on the phase space variables; we follow 
\cite{Louis-Martinez} and mainly \cite{Medved} where this issue is dealt with in detail 
for charged black holes arising in two dimensional dilatonic gravity. Let us 
nonetheless briefly summarize some key points and results: by direct
substitution of (\ref{laC}) and of the explicit expression for $\Pi_\rho$, $H_{can}$ in (\ref{Hamil}) becomes 
\begin{equation*}
H_{can}=\int\limits_\mathcal{C} dx\left[\frac{\dot{\phi}}{\phi}\mathcal{F}-\frac{1}{2}\left(
\frac{\sigma e^{2\rho}}{\phi^\prime}C^\prime -A_0 q^\prime\right)\right].
\end{equation*}
For any metric whose expression approaches asymptotically (\ref{metric})
in such a way that, at spatial infinity, $\dot{\phi}\to 0$ and $\frac{\sigma
e^{2\rho}}{\phi^\prime}\to 1$, the surface term can thus be chosen as 
\begin{equation*}
H_{sur.}=\int\limits_\mathcal{C} dx\frac{1}{2}\left(\frac{\sigma
e^{2\rho}}{\phi^\prime}C\right)^\prime+\left(A_0 q\right)^\prime,
\end{equation*}
where we can identify in the first term the ADM Hamiltonian $H_{ADM}$ and, 
consequently the ADM mass $M=\frac{C}{2}$.
In spite of this result, from now on, we will switch from the coordinates
$(C,q,\Pi_C,\Pi_q)$ for the phase space $\Gamma$ to the fully equivalent coordinates $(M,\Pi_M,q,\Pi_q)$.
Within this framework it is possible to write the action (\ref{act3}) as a
function on the reduced phase space i.e.
\begin{equation}\label{azioneiniziale}
I=\int dt\left[\Pi_M\dot M+\Pi_q\dot q- H(M,q)\right],
\end{equation}
where, according to the previous discussion, the Hamiltonian is independent from
the momenta. 

Let us now slightly specialize our analysis to a specific scenario where (\ref{metric}) describes a
black hole background and, thus, necessarily it exists at least a value for $\sigma$, say $\bar{\sigma}$,
such that $f(\bar{\sigma},M,q)=0$. Within this framework, the action (\ref{azioneiniziale}) represents 
the starting point in \cite{Barvinsky, Barvinsky2, Barvinsky3} to explicitly construct 
the black hole entropy spectrum and consequently the area spectrum since, as we
will emphasize later, we consider only black holes where the Bekenstein entropy to
area ratio holds. The first step in the analysis involves a Wick rotation
from the Lorentzian to the Euclidean metric. From a semiclassical
thermodynamical analysis, we can infer that the mass momentum becomes periodic i.e. $\Pi_M$ must be
identified with $\Pi_M+\frac{1}{T_H(M,q)}$ where $T_H(M,q)$ is the Hawking
temperature. Thus the phase space 
$\Gamma$ becomes topologically equivalent to $\mathbb{R}^3\times S^1$ and the circle, parametrized by $\Pi_M$, can be unwrapped
by means of a coordinate transformation $(M,\Pi_M,q,\Pi_q)\to
(X,\Pi_X,q,\Pi^\prime_q)$ :
\begin{gather}\label{prima}
\qquad \Pi_X=\sqrt{\frac{B(M,q)}{\pi }}\sin\left(2\pi\Pi_M
T_H(M,q)\right),\quad\Pi^\prime_q=\Pi^\prime_q(M,\Pi_M,q,\Pi_q),\\
\qquad\qquad\qquad\qquad X=\sqrt{\frac{B(M,q)}{\pi }}\cos\left(2\pi\Pi_M T_H(M,q)\right),
\label{seconda}
\end{gather} 
where both $B(M,q)$ and $\Pi^\prime_q(M,\Pi_M,q,\Pi_q)$ are functions which are
set by the request that the above change of coordinates is canonical i.e.
\begin{gather*}
\Pi_X\delta X+\Pi_q\delta q=\Pi_M\left(T_H(M,q)\frac{\partial B(M,q)}{\partial
M}\right)\delta M+\\
+\left(\Pi_q+\Pi_M T_H(M,q)\frac{\partial B(M,q)}{\partial q}\right)\delta q.
\end{gather*}
This implies:
\begin{equation*}
\frac{\partial B(M,q)}{\partial M}=T^{-1}_H(M,q),\quad\Pi^\prime_q=\Pi_q-T_H(M,
q)\frac{\partial B(M,q)}{\partial q},
\end{equation*}
where the first equality, together with the first law of thermodynamics for black
holes, grants us that $\frac{\partial B(M,q)}{\partial M}=\frac{\partial
S(M,q)}{\partial M}$ or, equivalently, upon integration
\begin{equation}\label{entrop}
B(M,q)=S(M,q)+F(q).
\end{equation}
The function $F(q)$ is an a priori arbitrary function though it is commonly
chosen as $-S_0(q)$, the minimum of the entropy $S(M,q)$ in terms of $M,q$. The
reasons behind this prescription are several, the most notable lying in the coordinate transformations
(\ref{prima}) and (\ref{seconda}) which grant us that the region $(M,\Pi_M)\sim
S^1\times\mathbb{R}\subset\Gamma$ is mapped into $(X,\Pi_X)\sim
\mathbb{R}^2-\mathbb{D}$ where $\mathbb{D}$ is a disk of radius
$S_0(q)+F(q)\geq 0$. Thus $F(q)=-S_0(q)$ grants us that the subspace $(X,\Pi_X)$
of the phase space with coordinates $(X,\Pi_X,q,\Pi^\prime_q)$ is 
$\mathbb{R}^2$. As shown in \cite{Barvinsky}, the above choice also uniquely
determine $\Pi^\prime_q$ as
\begin{equation*}
\Pi^\prime_q=\Pi_q+\Phi\Pi_M+\Pi_M T_H(M,q)\frac{d S_0(q)}{dq},
\end{equation*}
where $\Phi$ is the electrostatic potential at the boundary of the region of
spacetime under consideration.

Considering the phase space variables $(X,\Pi_X,q,\Pi^\prime_q)$ as observables on
a suitable Hilbert space of square integrable functions in such a way that they satisfy the canonical commutation relations
$[\hat{X},\hat{\Pi}_X]=[\hat{q},\hat{\Pi}_q]=i$, we can directly obtain the
area spectrum. By direct inspection of (\ref{prima}), (\ref{seconda}) and of
(\ref{entrop}) we end up with:
\begin{equation}
S(M,q)-S_0(q)=\pi \left(\hat{X}^2+\hat{\Pi}_X^2\right).
\end{equation}
The right hand side is a self-adjoint operator on the above mentioned Hilbert space
and it can be straightforwardly quantized as in ordinary quantum mechanics:
\begin{equation}\label{spectrum1}
S(M,q)-S_0(q)=2\pi \left(n+\frac{1}{2}\right).\qquad n\in\mathbb{N}
\end{equation}
In order to complete our task, we need also to quantize the electromagnetic
sector and this can be achieved following \cite{Barvinsky} where, starting from
(\ref{variazione}), it has been
shown that the phase space points $(q,\Pi^\prime_q)$ and
$(q,\Pi_q+2\pi\frac{n_1}{e}+n_2\frac{d S_0(q)}{dq})$
($n_1,n_2\in\mathbb{N}$) must be identified i.e. the subspace spanned by
$(q,\Pi^\prime_q)\sim S^1\times\mathbb{R}$. If we consider a momenta
representation for the charge operator i.e. $\hat{q}=i\frac{\partial}{\partial
\Pi_q}$, then the wavefunction for a charge eigenstate is proportional to 
$e^{iq\Pi_q}$ which must be single-valued.  Considering the above periodicity for the $q$-variable, the request of
singlevaluedeness implies that
\begin{equation*}
n_1\frac{q}{e}+n_2\frac{q}{2\pi }\frac{d S_0(q)}{dq}\in\mathbb{N},
\end{equation*}
which grants us the quantization rules:
\begin{equation}\label{spectrum2}
q=me,\quad \frac{q}{2\pi }\frac{d S_0(q)}{dq}=p.
\end{equation}
It is worthwhile underlining that the above equations are not independent and they 
ultimately provide a constraint on the possible values of $e$ in terms of $p$ and $m$ 
\cite{Barvinsky}. 
Thus, upon selecting a specific black hole background and consequently a specific
expression for $S_0(q)$, (\ref{spectrum1}) together with (\ref{spectrum2})
provides a quantized expression for the entropy. 
At this stage, some comments are in due course: 
\begin{itemize}
\item if we consider scenarios where the Bekenstein entropy to area ratio holds,
(\ref{spectrum1}), together with (\ref{spectrum2}), automatically provides an area spectrum,
\item $S_0(q)$ is the minimum of $S(M,q)$ and thus, if the relation 
$S(M,q)=\frac{A(M,q)}{4}$ holds, $A_0(q)$ represents the event horizon area for 
the black hole in the extremal configuration; a similar assertion cannot be 
stated for $S_0(q)$ since, in this specific scenario, such as for example the 
extremal Reissner-Nordstr\o m black hole, the classical entropy is vanishing and 
thus the Bekenstein formula does not seem to hold. This issue is nonetheless 
still under debate mainly due to apparent discrepancy between the results 
achieved from thermodynamical arguments and those inferred from a statistical 
counting of (extremal) black hole degrees of freedom in string theory. We 
refer to \cite{Kiefer} and references therein for a 
recent discussion on this specific problem.
\end{itemize}

\subsection{The spectrum of an asymptotically flat Reissner-Nord\\
str\o m black hole }

The approach discussed in the previous section can be applied to several 
scenarios though the most natural and one of the first analyzed is the 
asymptotically flat Reissner-Nordstr\o m black hole.
The starting point is (\ref{act1}) which appears to be a rather
specific subcase of (\ref{act3}) where 
the radius $r$ of the sphere plays the role of the
dilaton $\psi$ and where $V(r)=1$, $D(r)=\frac{r^2}{2}$ and
$W(r)=r^2$. Consequently, solving (\ref{conformalfactor}) as $\Omega^2(r)=r=
\sqrt{2\phi}$, we can recast (\ref{act1}) as
\begin{equation}\label{act2}
S=\int\limits_{\mathcal{M}^2} d^2x\sqrt{\left|\bar{g}\right|}\left[\frac{1}{2}\left(\phi
R(g_{\alpha\beta})+\sqrt{\frac{1}{2\phi}}\right)-\left(2\phi\right)^{\frac{3}{2}}
F^{(2)\alpha\beta}F^{(2)}_{\alpha\beta}\right].
\end{equation}
A solution for the Euler Lagrange equation for this action can be directly
inferred from (\ref{metric}), (\ref{metric2}) and (\ref{metric3}); since, in this
specific setting, $2\sigma=r^2$, we end up with $j(\sigma)=\sqrt{2\sigma}$ and
$k(\sigma)=-\left(2\sigma\right)^{-\frac{1}{2}}$ i.e., switching back from
$\sigma$ to the $r$ coordinate,
\begin{gather}
 F_{\alpha\beta}=-\frac{q}{r^2}\epsilon_{\alpha\beta},\\
 ds^2=-r\left[1-\frac{C}{r}+\frac{q^2}{r^2}\right]d\tau^2+r\left[1-\frac{C}{r}+
\frac{q^2}{r^2}\right]^{-1}dr^2,\label{RN}
\end{gather}
which, bearing in mind the conformal rescaling (\ref{coordch}) and plugging
(\ref{RN}) in (\ref{sym}), is the Reissner-Nordstr\o m metric with $C$ 
identified with twice the black hole mass $M$.
It is straightforward to realize, starting from (\ref{RN}), that the analysis in the previous section
can be applied to this specific scenario and thus we are entitled to construct the entropy/area spectrum. 
The unique step consists in remembering that the radii of the inner and outer 
event horizons are   
\begin{equation*}
r_\pm=M\pm\sqrt{M^2-q^2}.
\end{equation*}
Thus the entropy is  $S(M,q)=\pi r^2_+$ whereas the extremal configuration is 
achieved whenever $M=q$ (i.e. $r_+=r_-=q$) which implies either that $S_0(q)=\pi q^2$ either that (\ref{spectrum2}) becomes
\begin{equation*}
q=me,\quad q^2=p,
\end{equation*}
or equivalently $e^2=\frac{p}{m^2}$. This quantization rule for the charge can be set in (\ref{spectrum1}), eventually finding the area/entropy spectrum
\begin{equation}\label{RNspectrum}
S=\frac{A}{4G}=2\pi n+\pi(p+1).
\end{equation}
Here we have restored the explicit dependence on the Newton constant $G$ since,
bearing in mind that the adiabatic invariant is $A(M,q)-A_{ext.}(q)$, a
direct inspection of (\ref{RNspectrum}) shows that the value of $a_0$ in 
(\ref{Bek1}) can be set to $8\pi l^2_p$.
An interesting consequence of this result arises if we bear in mind that the
vacuum classical configuration of a Reissner-Nordstr\o m black hole corresponds 
to the extremal configuration with area $A_{ext.}=\pi  q^2=\pi  p$. If we interpret the 
quantum number $n$ in (\ref{RNspectrum}) as labelling the excited levels over 
the vacuum $n=0$, the classical lower bound is never reached due to the presence 
of an additional term $\pi$ which can be interpreted as a symptom of vacuum 
fluctuations. Thus we can conclude that the extremal configuration does not lie 
in the spectrum constructed with this specific quantization technique and this 
feature will be common also when we will deal, in the next section, with a non 
vanishing cosmological constant scenario.
 
\section{The Reissner-Nordstr\o m AdS black hole}
Let us now address the main question of this paper namely if we can extend the
above results for a charged non rotating asymptotically flat black hole to the 
asymptotically AdS counterpart. Adopting the same notations and conventions from 
the previous sections, the starting point is the Einstein-Maxwell action with a
cosmological term i.e. \cite{Wald}
\begin{equation}\label{ADSact}
S=\int\limits_{\mathcal{M}^4} d^4x \sqrt{\left|g\right|}\left(\frac{R-2\Lambda}{16\pi}-\frac{
F_{\mu\nu}F^{\mu\nu}}{4}\right),
\end{equation}
where $\Lambda<0$ is the cosmological constant. If we look for spherically
symmetric solutions for the Euler-Lagrange equations, we can introduce a
coordinate frame $(x_0,x_1,\theta,\varphi)$ and we can write the metric as
\begin{equation*}
ds^2=g_{\alpha\beta}dx^\alpha dx^\beta+r^2(x_\alpha)d\Omega^2(\theta,\varphi),
\end{equation*}
where $g_{\alpha\beta}$ is a two dimensional metric
depending upon the coordinates $x_\alpha\doteq\left\{x_0,x_1\right\}$ spanning a two
dimensional submanifold $\mathcal{M}^2\subset\mathcal{M}^4$.\\
Assuming that also the $U(1)$ vector potential $A_\mu$ is spherically symmetric,
we look for the dimensionally reduced action plugging the above metric in 
(\ref{ADSact}). A straightforward calculation ends up with
\begin{eqnarray}
S=\negmedspace\negmedspace\int\limits_{\mathcal{M}^2} d^2x\sqrt{\left|g\right|}\left[
\left(\frac{g_{\alpha
\beta}}{4}\partial^\alpha r\partial^\beta
r+\frac{1}{2}+\frac{r^2}{4}\left(R(g_{\alpha\beta})-2\Lambda\right)\right)-\frac{r^2}{4}F^{(2)\alpha\beta}
F^{(2)}_{\alpha\beta}\right],\label{act6}
\end{eqnarray}
This is again a special case of (\ref{act3}) with $r(x_\alpha)$ playing the role
of the dilaton field, $D(r)=\frac{r^2}{2}$, $W(r)=r^2$, but $V(r)=1-\Lambda
r^2$. Thus we can exploit (\ref{coordch}) and (\ref{conformalfactor}) redefining
the dilaton as $\phi=\frac{r^2}{2}$ and introducing the conformal factor
$\Omega^2(\phi)=r=\sqrt{2\phi}$; consequently the above action is transformed in
\begin{eqnarray}
S=\int\limits_{\mathcal{M}^2} d^2x\sqrt{\left|\bar{g}\right|}\left[\frac{1}{2}\left(\phi
R(g_{\alpha\beta})+\sqrt{\frac{1}{2\phi}}-\Lambda\sqrt{2\phi}\right)-\left(2\phi\right)^{\frac{3}{2}}
F^{(2)\alpha\beta}F^{(2)}_{\alpha\beta}\right].\label{act5}
\end{eqnarray} 
A solution for the Euler-Lagrange equation for this action can be inferred
directly from (\ref{metric}), (\ref{metric2}) and (\ref{metric3}), namely,
bearing in mind that the role of the spatial coordinate $\sigma$ is played by
$\phi$, we calculate $k(\sigma)=-(2\sigma)^{-\frac{1}{2}}$ whereas
$j(\sigma)=\sqrt{2\sigma}-\frac{\Lambda}{3}(2\sigma)^{\frac{3}{2}}$. Switching
back to the $r$-coordinate by means of $r^2=2\sigma$, we end up with
\begin{gather}
F_{\alpha\beta}=-\frac{q}{r^2}\epsilon_{\alpha\beta},\\
\qquad ds^2=-r\left[1-\frac{C}{r}+\frac{q^2}{r^2}-\frac{\Lambda}{3}r^2\right]d\tau^2+r
\left[1-\frac{C}{r}+\frac{q^2}{r^2}-\frac{\Lambda}{3}r^2\right]^{-1}dr^2,
\end{gather}
which is, up to the conformal rescaling (\ref{coordch}),
the AdS Reissner-Nordstr\o m metric with $C$ identified with twice the black
hole mass $M$ (see \cite{Carter} and sections 24.2, 24.4 in \cite{Kramer}).

As it is shown in the previous discussion, the main effect of the cosmological
constant from the point of view of dimensionally reduced gravity concerns the
variation of the potential $V(r)$ which is no more a constant, but it acquires a
term directly dependant on the $r$-variable. Nonetheless, from the point of view
of the Hamiltonian approach to black hole spectroscopy, we can still repeat from scratch the analysis
discussed in section \ref{II}. Thus we can introduce a generic two dimensional metric
$ds^2=e^{2\rho}\left[-N_1 dt^2+ (dx+N_2dt)^2\right]$ and we can  repeat the
same analysis ending up with the Hamiltonian (\ref{Hamil}).\\
The only subtlety lies in the discussion of the boundary term $H_{surf}$;
within this respect it is interesting to notice that the analysis in 
\cite{Louis-Martinez}, summarized in section \ref{II}, do not require a priori 
the asymptotical flatness of the metric, but only that it tends at least 
asymptotically to (\ref{metric}), thus it can be applied to an AdS background. We 
will not discuss this issue in details in this paper; 
suffice to say that a more detailed analysis of the Hamiltonian dynamic for a 
charged AdS black hole has been presented in \cite{Louko} ({\em c.f.} sections II e 
IV) though without the language of dilatonic gravity. In particular it is shown 
that, also for an AdS Reissner-Nordstr\o m black hole, it is possible to 
construct a reduced phase space depending only upon the black hole mass $M$, the 
$U(1)$ electric charge $q$ and the conjugate momenta respectively $\Pi_M$ and 
$\Pi_q$. We Wick-rotate the Lorentzian time to the Euclidean counterpart in such
a way that, in order to avoid a conical singularity in the metric, the latter 
must be periodic with period $T^{-1}_H(M,q)=4\pi r_+\left(1-\frac{q^2}{r^2_+}-
\frac{\Lambda r^2_+}{2}\right)^{-1}$, being $r_+$ the outer horizon radius of the
AdS Reissner-Nordstr\o m black hole. The same result translates to $\Pi_M$
i.e. $\Pi_M\sim\Pi_M+T^{-1}_H(M,q)$; slavishly repeating the construction from the
previous section, it is now possible to demonstrate that the Hamiltonian 
ultimately depends only upon the four mentioned reduced phase space variables 
and on the boundary conditions, thus it can be written as (\ref{azioneiniziale}).

Eventually we are also entitled to consider in the reduced phase space $\Gamma$ 
the canonical transformations of variables
(\ref{prima}) and (\ref{seconda}). The final result mimics that of an
asymptotically flat Reissner-Nordstr\o m black hole i.e.
\begin{equation*}
S(M,q)-S_0(q)=\pi\left(\hat{X}^2+\hat{\Pi}^2_X\right),
\end{equation*}
where we have interpreted the phase space variables as operators on a suitable
Hilbert space. Consequently we end up with the following entropy spectrum:
\begin{equation}\label{spectrumads1}
S(M,q)-S_0(q)=2\pi\left(n+\frac{1}{2}\right).
\end{equation}
The real difference in comparison with the scenario considered in the previous section lies in
the expression for $S_0(q)$ and in the quantization for the electromagnetic
sector. As discussed previously, $S_0(q)$ stands, classically, for the entropy 
of an extremal black hole i.e.
\begin{equation}\label{extremal}
S_0(q)=\pi r^2_{ext}=\pi \left[\frac{1-\sqrt{1-4q^2\Lambda}}{2\Lambda}\right],
\end{equation}
whereas the charge can be quantized by means of (\ref{spectrum2}) i.e.
\begin{equation}\label{AdScharge}
q=me,\quad\frac{q^2}{\sqrt{1-4q^2\Lambda}}=p.
\end{equation}
As stated in section \ref{II}, these two quantization conditions must be simultaneously satisfied providing a constraint for the value of $e$ namely
\begin{equation*}
e^2=\frac{-2m^2p^2\Lambda+\sqrt{8m^4\Lambda^2 p^4+p^2m^4}}{m^4}.
\end{equation*}
By direct substitution of (\ref{AdScharge}) in (\ref{extremal}), we find
\begin{equation*}
S_0(q)=\frac{\pi}{2\Lambda}+\pi p\left(1+\sqrt{1+\frac{1}{4\Lambda^2 p^2}}\right),
\end{equation*}
which, together with (\ref{spectrumads1}), provides the full AdS Reissner-Nordstr\o m spectrum:
\begin{equation}\label{spectrumads2}
S(M,q)=2\pi n+\pi p\left(1+\sqrt{1+\frac{1}{4\Lambda^2 p^2}}\right)+\pi\left(1+\frac{1}{2\Lambda}\right).
\end{equation}
At this stage, a few remarks are in due course:
\begin{itemize}
\item The levels in (\ref{spectrumads2}) are not equispaced though it appears that, 
whenever $\Lambda p\gg 1$, the entropy spectrum can be approximated with a formula granting equispaced levels.
This limit can be achieved in two ways; in the first we consider great integer values for the quantum 
number $p$ whereas, in the second case, we deal with the limit $\Lambda\gg 1$. 
In this latter scenario, it is interesting to remark that the spectrum tends to
its asymptotically flat counterpart (\ref{RNspectrum}).
\item the spectrum (\ref{spectrumads2}) is consistent with the AdS-Schwarzschild black hole namely if we consider a vanishing electric charge, the electric sector contribution cancels i.e. 
$S(M)=2\pi n + \pi$. This is a universal result proper of any two dimensional dilatonic gravity and thus also of the spherically symmetric reduced scenario of a Schwarzschild black hole, 
independently from the value of the cosmological constant \cite{Barvinsky4,
Barvinsky5}.
\item if we interpret the quantum number $n$ as a measure of the excitation of 
the AdS Reissner-Nordstr\o m black hole above extremality, we find, setting 
$n=0$ in (\ref{spectrumads2}) and independently from the charged sector, a non 
zero entropy which can be interpreted 
as the symptom of vacuum fluctuations. This result coincide with that achieved 
in the asymptotically flat counterpart, namely, within this quantization scheme 
the extremal configuration does not lie in the spectrum. To a certain extent 
this result confirms the statement (see \cite{Barvinsky3} and \cite{Das3}) according to 
which a third law of thermodynamic for black holes can be formulated: it is 
impossible that a non extremal black hole decays in an extremal configuration 
in a finite sequence of physical processes.
\item bearing in mind that the adiabatic invariant for a non neutral black hole
is the difference between the area and its extremal value, we can infer from a
direct inspection of (\ref{spectrumads1}) the value of $a_0$ in (\ref{Bek1}) i.e.
$8\pi l^2_p$ as in the asymptotically flat Reissner-Nordstr\o m scenario.
The reader should also take into account that the constant term in (\ref{spectrumads1})
was not considered in Bekenstein ansatz though it is a natural feature both in
the reduced phase space approach and in the axiomatic one discussed in the
appendix.
\item within the framework of Bekenstein formulation of black hole spectroscopy, a
key result lies in the equivalence for a charged black hole between the reduced
phase space approach and the axiomatic one, the latter reviewed in the appendix. 
As shown in \cite{Das} it is possible to start from the Hamiltonian formulation
on the reduced phase space and to construct, as in the axiomatic scenario, an algebra of operators in terms of the variables
$(M,\Pi_M,Q,\Pi_Q)$. Within this framework the two above mentioned approaches are identical
with a due exception: the area operator considered by Bekenstein is equivalent in the reduced
phase space approach to the the difference between the area of the event horizon and the area
in the extremal configuration. Furthermore it appears that the eigenvalue which 
is obtained in \cite{Das} i.e. (\ref{spectrumads1}) is different from the Bekenstein 
one (\ref{axiomeig2}) up to a sign in the lowest
eigenvalue; to the best of our knowledge such discrepancy has no clear explanation and it should
be investigated carefully. Nonetheless the reasoning of \cite{Das} 
can be smoothly extended to our scenario and also, in an extremal AdS Reissner-Nordstr\o m black hole, the
equivalence between the axiomatic and the reduced phase space methods can be 
claimed.
\end{itemize}
\section{Conclusions}
In this paper we have explicitly constructed the area spectrum (\ref{spectrumads2})
for an AdS Reissner-Nordstr\o m black hole and, in particular, we have shown that the 
contribution from the extremal part and the charged sector breaks the equispacing
between the eigenvalues of the area operator. This result is ultimately true
only in the Euclidean regime as in the asymptotically flat scenario, but we are
not aware of a universally accepted derivation of area spectra in a Lorentzian
background.
Nonetheless we believe that the result could be highly interesting if considered as a
starting point for future line of researches. In particular, besides addressing the
natural question of the de Sitter counterpart of our analysis, the next direct step
consists of considering a rotating black hole. In the
asymptotically flat case, the Kerr-Newman solution has been studied in detail in
\cite{Gour}, the main a priori difficulty lying in the non separability of the
charge and the spin sector. This implies that this scenario cannot be seen as a simple
extension of the charged non rotating case \cite{Barvinsky} or of the neutral
rotating case \cite{Gour2}; a similar consideration could be applied to the results of
this paper and thus the analysis of the AdS-Kerr-Newman black hole deserves careful
considerations.

A further interesting application of the reduced phase space approach to black hole spectroscopy
has been developed in section V of \cite{Das4}, where it has been shown 
that a direct consequence of an equispaced area/entropy spectrum is a markedly discrete
spectrum for the Hawking radiation in the considered background. On the opposite, when 
the area spectrum in not equispaced (but discrete) such as in the three dimensional BTZ 
black hole or in the five dimensional rotating black hole, the corresponding spectrum for
the Hawking radiation is quasi-continuum. Thus, in our scenario, we can slavishly repeat the analysis 
from the above cited paper starting from the first law of thermodynamics for a charged non rotating black 
hole
$$\delta M=T_H\delta S+ \phi\delta Q,$$
where $T_H$ is the Hawking temperature and $\phi$ is the electrostatic potential which on the horizon can
be set to $\phi=\frac{Q}{r_+}$.
If we take into account that $\left|\delta M\right|$ is proportional to the lowest frequency $\omega_0$
emitted  in the Hawking radiation for the transition of a black hole from an excited level to its next 
lower state, the above equation becomes
\begin{equation}\label{Hawk}
\delta\omega_0=\pi T_H\left[\delta n+\left(1+\frac{1}{2\Lambda p\sqrt{1+4\Lambda^2 p^2}}\right)\delta p
\right]+\frac{e^2 p}{r_+}\delta p,
\end{equation}
where we have set $Q=pe$ and we have used the entropy spectrum formula (\ref{spectrumads2}). 
A direct inspection shows that the AdS Reissner-Nordstr\"om black hole shares the same feature than the black
hole considered in \cite{Das4} i.e. if we consider a variation $\delta n=1$ but $\delta p=0$, 
the spectrum appears to be equispaced. This is not surprising since it is a direct consequence of the reduced
phase space approach which ultimately grants us that the spectrum of the black hole entropy minus the 
contribution from the extremal configuration is that of an harmonic oscillator. Conversely, if we
choose $\delta n=0$ but $\delta p=1$, the spectrum is not equispaced and if we consider macroscopic 
black holes and large quantum numbers (i.e. $n,p\gg 1$ and $r_+\gg e^2p$), (\ref{Hawk}) tends to $0$
for $p\to\infty$ i.e. the spectrum for the Hawking radiation appears to be quasi-continuum. It is fair to 
admit that this is more a preliminary remark more than a definitive claim; hence it would be interesting
to repeat the above analysis in different asymptotically AdS backgrounds since the physical origin of
the above behaviour apparently lies in the presence of the 
cosmological constant\footnote{In the AdS-Schwarzschild scenario, the spectrum for the Hawking radiation is clearly
equispaced since the only contribution to the entropy comes from the ``harmonic oscillator'' as we have 
underlined at the end of the previous section}.  

On a broader setting, as briefly stated in the introduction, the results from 
this paper should be compared with different methods of area quantization the most 
notables
arising from loop quantum gravity \cite{Rovelli} and from the interpretation of 
the discreteness of the spectrum as an absorption of quasi normal mode 
excitations \cite{Dreyer, Hod} (see also \cite{Setare} for an application to
asymptotically flat extremal Reissner-Nordstr\o m black holes and \cite{Setare2} for an
application to BTZ black holes). In particular the area spectrum calculated in this 
last approach is completely different from the one calculated with the Hamiltonian
framework we have introduced in the previous sections. In the asymptotically flat
scenario there is no deep explanation for this discrepancy and, thus, we believe that
the results in this paper could be, from the Hamiltonian point view, a starting point 
to extend the comparison between the above two mentioned methods in a wider class of 
spacetimes though the analysis of area spectra from quasi-normal modes in an AdS 
background is still far from complete (for a calculation of quasi normal modes for the
AdS Reissner-Nordstr\o m black hole see \cite{Wang}).

Moreover it would be of great interest to analyze the results in this paper from the point of view of an
holographic reconstruction of black hole physics. This specific point of view has not
been pursued in the previous analysis of black holes spectra mainly due to the lack of
a complete construction of an holographic correspondence in an asymptotically flat
background. Conversely, in this paper we dealt with an asymptotically AdS spacetime
which is the natural setting for the (asymptotically) AdS/CFT correspondence and it is important 
to remember that the black hole entropy is related to a conformal
group or, more properly, to the associated Virasoro algebra by means of the Cardy formula. 
The usual rationale consists on constructing a conformal group on the horizon itself though it
has been shown that this is not a unique point of view and it is also possible to deduce 
the statistical black hole entropy starting from the conformal asymptotic symmetry group as 
in \cite{Cadoni} where the entropy of a two dimensional black hole\footnote{The reader
should bear in mind that, as it was discussed in the previous sections, a four dimensional spherically symmetric 
black hole can be equivalently described by means of a two dimensional black hole and of a a dilaton field 
by means of a dimensional reduction procedure.} is 
calculated in the framework of the $AdS_2/CFT_1$ correspondence. Thus, from a broader point of view,
our hope is that the area operator for an AdS black hole could always be interpreted as a 
suitable dual quantum mechanical operator in the boundary conformal field theory and that the analysis 
of the spectrum of such an operator could be the same or at least related with the 
one proposed in our analysis by means of a semiclassical argument. To the best of our
knowledge, such line of research has not been pursued yet and we are currently 
investigating it.

\appendix
\section{The axiomatic approach}\label{I}
The first approach to black hole area quantization, often referred to as the 
{\em algebraic or axiomatic formulation},
has been discussed in \cite{Bekenstein4, Bekenstein5, Bekenstein6, Gour3}. The
underlying rationale consists of introducing a suitable set of operators
defining a closed algebra by means of their commutation relations. The starting
point lies in the intuitive assumption that it exists a suitable Fock space
$\mathcal{F}$
whose elements $\left| nqjms\rangle\right.$ describe a one-black hole state which is
characterized by the eigenvalues of the following set of operators ($\hbar=1$):
\begin{itemize}
\item an area operator $\hat A$ such that $\hat A\left|
nqjms\rangle\right.=a_n(q,j,m)\left| nqjms\rangle\right.$,
$n\in\mathbb{N}$, $a_0\doteq 0$ and $a_{n+1}>a_n$,
\item a charge operator $\hat q$ such that $\hat q \left|
nqjms\rangle\right.=qe\left| nqjms\rangle\right.$ where $q\in\mathbb{Z}$.
\item the usual angular momentum operators $\hat{J}^2, \hat{J}_{e_i}$ such that 
$\hat{J}^2\left|
nqjms\rangle\right.=j(j+1)\left| nqjms\rangle\right.$ with $j$ a positive
integer or semi-integer. $\hat{J}_{e_i}$ is the projection of $\hat{J}$ along a fixed
direction $e_i$ and $\hat{J}_{e_i}\left| nqjms\rangle\right.=m\left|
nqjms\rangle\right.$ with $m=-j,....,j$ for a fixed $j$.
\end{itemize}
The last quantum number, $s$, distinguishes between elements in $\mathcal{F}$
with the same values of $(nqjm)$ and thus it runs from $1$ to the degeneracy
$g_n$ of the state $\left| nqjms\rangle\right.$. It is possible to
further characterize a state with fixed value of $s$ introducing a suitable
operator which admits $s$ itself as an eigenvalue \cite{Gour}. Unfortunately such
approach requires that the equispacing of the spectrum of the area operator 
is a priori imposed, but this is a strong request which may not be satisfied in
several physically interesting scenarios. Conversely, we will stick to the approach, first
introduced in \cite{Bekenstein4} and in \cite{Bekenstein6}, where it is assumed 
that it exists an operator $\hat{R}_{nqjms}$ such that 
\begin{equation*}
\hat{R}_{nqjms}\left|vac\rangle\right. =\left| nqjms\rangle\right.,
\end{equation*}
where $\left|
vac\rangle\right.\in\mathcal{F}$ is the unique state $\left|00000\rangle\right.$.\\
If we also introduce the identity $\hat{I}$, the set of operators
$\hat{A},\hat{q},\hat{J}^2,\hat{J}_{e_i},\hat I$ 
together with $\hat{R}_{nqjms}$ is required to form a \emph{closed and linear} 
algebra; besides the natural property of closure, the linearity is a strong
assumption justified only by the physical prescription that $\hat{A}$ forms an
additive quantity whenever several black holes are taken into account. Following
the calculations of \cite{Bekenstein4, Bekenstein5}, we introduce for simplicity the usual 
operators $\hat{J}_\pm=\hat{J}_{e_l}\pm i\hat{J}_{e_k}$ where 
$(e_i,e_k,e_l) $  form an orthonormal base in $\mathbb{R}^3$ and where 
$\hat{J}^2=\frac{1}{2}\left(\hat{J}_+\hat{J}_-+\hat{J}_-\hat{J}_+\right)+
\hat{J}^2_{e_i}$; the algebra defining commutation relations are:
\begin{gather}
\left[\hat q,\hat{R}_{nqjms}\right]=qe\hat{R}_{nqjms},\quad \left[\hat J_{e_i},
\hat{R}_{nqjms}\right]=m\hat{R}_{nqjms},\\
\left[\hat J_{\pm},
\hat{R}_{nqjms}\right]=\sqrt{j(j+1)-m(m\pm q)}\hat{R}_{nqjm\pm 1s},\\
\left[\hat A, \hat{R}_{nqjms}\right]=a_n
\hat{R}_{nqjms}+\delta^0_q\left[\delta^0_j\left(D_{ns}\hat{q}+E_{ns}\hat{A}
\right)+\delta^1_jF_{ns}\hat{J}_{e_i}\right],
\end{gather} 
where $D_{ns},E_{ns},F_{ns}$ are arbitrary complex numbers for all $n,s$ and 
where all other commutators either are $0$ or they assume the canonical expression as in quantum mechanics.

Starting from the above ansatz, it is possible to construct the area spectrum
for a spherically symmetric black hole as in \cite{Bekenstein4} and as in 
\cite{Bekenstein5}. Without entering into the details of Bekenstein calculations,
let us nonetheless  show the main results. 
The first step consists of noticing that the above 
commutation relations, in particular $[\hat{A}, J_{e_i}]=0$, grant us the rotational 
invariance for the area eigenvalues i.e. $a_n(q,j,m)=a_n(q,j)$. Moreover 
a similar argument excludes 
the dependence on the $j$ quantum number and, eventually, the one-black hole area 
operator depends only on the charge $q$; thus a long and tedious calculation
shows that the eigenvalues $a_n$ must satisfy, 
for a neutral black hole, the identity
\begin{equation}\label{axiomeig}
a_n(0)=na_1(0),
\end{equation}
whereas, for a charged black hole, 
\begin{equation}\label{axiomeig2}
a_n(q)=\left(n-\frac{1}{2}\right) a_1(0).
\end{equation}
Thus $a_1(0)$ plays the role of $a_0$ in (\ref{Bek1}) and, whenever the black
hole is charged, (\ref{axiomeig2}) shows that there is also a constant
contribution which was not present in (\ref{Bek1}).
Let us conclude with two important remarks: from one side the above discussed approach does not
provide any means to explicitly calculate a precise value for $a_1(0)$; from the
other side either (\ref{axiomeig}) either (\ref{axiomeig2}) grant us that
the area spectrum for a black hole is uniformly spaced without any need to
assume a priori such behaviour as in \cite{Gour}.

\section*{Acknowledgments}

The authors wish to thank Mauro Carfora for useful discussions and the referee for pointing out the relation
between Hawking radiation and black hole spectroscopy as developed in Ref.~\cite{Das4}. The work of C.D. is supported by a grant 
from the Department of Theoretical and Nuclear Physics - University of Pavia.


\end{document}